\documentclass[twocolumn,showpacs,preprintnumbers,amsmath,amssymb]{revtex4-1}
\usepackage{hyperref}
\pdfoutput=1

\usepackage{graphicx}
\usepackage{dcolumn}
\usepackage{amsfonts,amsmath,amssymb,bm}
\usepackage{float}
\usepackage{xcolor}

\begin{document}


\title{\bf {Direct band gap in gallium sulfide nanostructures}}
\date{\today}
\author{M.~Mosaferi}
\author{I.~Abdolhosseini Sarsari}
\author{M.~Alaei}
\affiliation{Department of Physics, Isfahan University of 
Technology, Isfahan, 84156-83111, Iran}

\begin{abstract}

The monolayer Gallium sulfide (GaS) was demonstrated as a promising two-dimensional 
semiconductor material with considerable band gaps. The present work investigates 
the band gap modulation of GaS monolayer under biaxial or uniaxial strain by using 
Density functional theory calculation. We found that GaS monolayer shows an indirect 
band gap that limits its optical applications. The results show that GaS monolayer has a 
sizable band gap. 
The uniaxial strain shifts band gap from indirect to direct in Gallium monochalcogenides (GaS).
This behavior, allowing 
applications such as electroluminescent devices and laser. The detailed reasons for 
the band gap modulation are also discussed by analyzing the projected density of states (PDOS). 
It indicates that due to the role of p$_y$ orbital through uniaxial strain become more significant 
than others near the Fermi level. The indirect to direct band gap transition happen at $\varepsilon$=-10y$\%$. 
Moreover, by investigating the strain energy and transverse response of structures under uniaxial strain, 
we show that the GaS monolayer has the Poisson's ratio of 0.23 and 0.24 in the zigzag (x) and armchair (y) 
directions, respectively. Thus, we conclude that the isotropic nature of mechanical properties under strain. 

\end{abstract}

\maketitle

\section{INTRODUCTION}

The 2D materials hold great promise in the design of new nano-devices for future applications, due to their fascinating electronic, mechanical, 
optical, and thermal properties which are caused by their low dimensionality and quantum confinement effect. 
The some of these novel properties not seen in their bulk counterparts \cite{gupta2015recent}. 2D nano-sheets 
with hexagonal structure, such as graphene, hexagonal boron nitride (h-BN) and transition metal 
dichalcogenides (TMDs) have attracted extensive research effort in recent years.
The utility of these materials, however, is limited by some imperfections such as the lack of a band-gap in
graphene and the relatively low mobility in some TMDs \cite{radisavljevic2011single}. It has motivated continuing work in search of 
more 2D materials that demonstrate their properties that may lead to improving specific performance.

\begin{figure}[!ht]
\centering
\includegraphics*[scale=16.5]{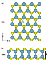}
\caption
{A schematic showing the crystal structure of GaS monolayer with (a) top view and (b) side view. 
The blue and yellow spheres represent the Ga and S atomic species, repectively. $a_0$ and $d$ 
represent the lattice constant and the thickness of the GaS monolayer as defined by the distance 
between the top and bottom layer of sulfur atom. The x and y orientations correspond
 to the zigzag and armchair direction, respectively.}
 \label{GaS-structure}
\end{figure}

Recently, another class of 2D materials, metal chalcogenides, came into being and attracted considerable attention. 
These layered materials generally possess the chemical formula MX, where M= Group IIIA and X= Group IVA. 
Gallium sulfide is an indirect band gap semiconductor.
 A primitive cell of GaS contains four Gallium atoms and four sulfur atoms.
 Its structure is formed from four layered of S-Ga-Ga-S type as illustrated in Fig.~\ref{GaS-structure}. 
The interlayers dominantly experience  the weak Van der Waals interaction, while within the layer the Ga and X atoms 
form the strong covalent bond similar to graphene and TMDs.

 GaS nanosheet was successfully isolated by micromechanical cleavage technique \cite{late2012rapid}\cite{aono1993near}. 
 Monolayer gallium sulfide has been used in photodetectors, photoelectric devices, electrical sensors, 
 near-blue light emitting devices \cite{shen2009vapor}\cite{hu2013highly}\cite{aono1993near},
and field effect transistors (FET). The mobility of 0.1 $cm^2/V.S$ have been measured 
by David et al \cite{late2012gas} in ultra-thin GaS bottom-gate transistors. 
Moreover, the piezoelectric coefficients of GaS monolayer are of the same order 
of magnitude as discovered two-dimensional (2D) piezoelectric materials. 
(The boron nitride 
(BN) and $MoS_2$ monolayers \cite{li2015piezoelectricity}).

\begin{figure}[!ht]
\centering
\includegraphics*[scale=28]{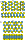}
\caption{Top and side views of the bare-edge (a) armchair GaS nanoribbon with $N_a$=6 and (b) zigzag GaS nanoribbon with $N_z$=8. The width $N_a$ and $N_z$ of AGaSNR and ZGaSNR are
depicted by black dashed line. The blue and yellow spheres represent Ga and S atoms, respectively.}
\label{nanoribbon-stracture}
\end{figure}

The indirect band gap of GaS monolayer presents an obstacle for optoelectronic devices applications involving
light harvesting, light emitting, photodetectors and lasers. The thicker thinfilm with direct optical band gap is desired. 
The strain plays an important role to engineer band gap of semiconductors in the field of microelectronic. 
It would be very interesting to investigate the effect of strain on the band structure of GaS. 
The effect of strain on graphene and TMDs have been widely studied\cite{topsakal2010response}\cite{choi2010effects}\cite{peelaers2012effects}\cite{scalise2012strain}
\cite{ghorbani2013strain}\cite{sun2016indirect}\cite{wang2015strain}.
 The band structure of these can be remarkably
tailored by applying strain \cite{ni2008uniaxial}\cite{huang2011strain}.
If the energy difference of the indirect and direct band gap is small, then it may be possible to achieve 
indirect-direct band gap transition. 
The GaS nanoribbons can be directly obtained by cutting the GaS monolayer (Fig.~\ref{nanoribbon-stracture}). 
Depending on the direction of termination, there exist two kinds of nanoribbons: armchair and zigzag. The armchair GaS
nanoribbons and zigzag GaS nanoribbons can be identified by the number of dimer lines or zigzag chains across the 
ribbon width and are labeled as $N_a$-AGaSNR and $N_z$-ZGaSNR, respectively.

In the present work, we have studied the electronic properties of GaS monolayer under strain. The strain 
can easily perform by a lattice mismatch on the substrate or mechanical loading\cite{shi2013quasiparticle}. 
we employed two type of deformations: uniaxial and biaxial. The uniaxial strain is parallel to x and y directions
in the range of $\varepsilon$=-10$\%$ to $\varepsilon$=+10$\%$.  
The biaxial compressive strain can firstly increase the band gap, up until 2.76 eV at $\varepsilon$=-4$\%$ and then reduce it, 
while biaxial tensile strain reduces the band gap gradually. Both of them can not change the indirect band gap to direct band gap.
Monolayer GaS undergoes indirect to direct band gap transition under 
uniaxial compressive strain along y-direction. It can indicate that as biaxial strain increasing the semiconductor to metal transition 
occurs by the energy crossover of one state in conduction band from Fermi level. Also, we focused on the
effect of edge passivation on the electronic band structures. 
We demonstrated that the electronic properties
of edge zigzag and armchair GaS nanoribbons can be largely tuned by edge passivation. Therefore, offers a 
powerful tool for customizing to a particular application. 
The band gap of nanoribbons can be also reduced or enlarged, depending on the direction 
of termination (zigzag or armchair) and width of nanoribbons.


\section{METHOD}
The calculations were performed using ab initio density functional theory (DFT) in conjunction with 
projector augmented wave (PAW) potentials \cite{blochl1994projector} and Perdew-Burke Ernzerted (PBE) generalized
gradient approximation (GGA) to the electronic exchange and correlation \cite{perdew1996jp}, as implemented 
in the Quantum Espresso (QE) package \cite{giannozzi2009quantum}.
The K-point mesh in the lateral directions was always 9$\times$9$\times$1 while,
the 8$\times$1$\times$1 grid K-points was used for armchair nanoribbons 
and 1$\times$12$\times$1 for zigzag nanoribbons. 
A kinetic energy cutoff of 500 eV was used for the plane-wave expansion.
Test calculation with the bigger number of K-points and higher cutoff gave essentially the same results. Bader analysis was used to
calculate charge transfer between Ga and S under different strain by Critic2 \cite{henkelman2006fast}.\\
The layered GaS is modeled using a supercell method with 2D periodic boundary conditions. 
A vacuum region of 20 {\AA} along the c direction was used to separate the layers system in order to
avoid spurious interactions. The geometric 
structures of the GaS monolayer are optimized by the conjugate-gradient minimization 
scheme. All of the atoms in the unit-cell are fully relaxed until the force on each atom is minimized. 
Our calculated lattice
constant for GaS monolayer is 3.64 {\AA}, and the band gap of 2.45 eV in good 
agreement with other theoretical calculation \cite{zolyomi2013band}\cite{huang2015effects}.

\section{RESULTS AND DISCUSSIONS}
\begin{figure*}[ht]
\centering
\includegraphics*[scale=12]{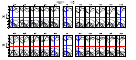}
\caption{The band structure calculated under tensile and compressive strain along 
(a) zigzag and (b) armchair directions. The variation of states A, B and C with strain are marked out by 
dashed lines.}
\label{band-uniaxial}
\end{figure*}

Here, we examine the band gap evolution under biaxial and uniaxial strain. 
In spite of DFT band gap underestimation in semiconductor, the Si 
nanostructures strain-dependency is similar compared to optical 
band gap predicted by the quasiparticle and advance configuration interaction methods \cite{peng2006strain}.
Therefore, we can predict the general trend of the strain effect on band structure.
The present work is mainly focused on the strain effect on the electronic band structure by the GGA-PBE. 
As expected not only the Ga-S bond length but also the distance between S atoms vary with applied strain, thus, 
the band gap can be significantly tuned by biaxial and uniaxial strain as shown in Fig.~\ref{band-biaxial} and Fig.~\ref{band-uniaxial}. 
So the GaS monolayer with an indirect band gap can tune into a direct band gap with a suitable
uniaxial strain and experiences a semiconductor to metal transition with biaxial tensile strain.\\

 The equilibrium structure was exposed to mechanical strain via two strategies: 
First, biaxial deformation, 2D isotropic deformation on the hexagonal unit cell with simultaneous change of $a_0$ and $b_0$ 
lattice vector. 
Second, uniaxial deformation of $a_0$ ($b_0$) along the zigzag (armchair) direction x (y) in the rectangular unit cell. 
The strain along the x and y direction are given by 

\begin{align}
\epsilon_x=(a_x-a_0)/a_0 
\end{align}
and
\begin{align}
\epsilon_y=(b_y-b_0)/b_0
\end{align}
respectively. Where $a_x$ and $b_y$ are deformed lattice vector along x and y directions. The calculated lattice parameters 
$a_0$ and $b_0$ in the unstrained structure of GaS monolayer are 3.64 {\AA} and 2.35 {\AA}, respectively. 

\subsection{Biaxial strain}
 
\begin{figure*}[ht]
\centering
\includegraphics*[scale=14]{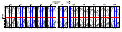}
\caption{First-principles band structure of monolayer GaS under biaxial tensile and compressive strain. 
The negative value indicates the compressive strain. The VBM and CBM are labeled by
A and B. The dashed lines are guide for an eye for energy shifts of states. 
The horizontal dashed lines indicate the Fermi level. 
The fermi level is set to zero.}
\label{band-biaxial}
\end{figure*}

 Our DFT calculations found that the strain-free GaS monolayer is the semiconductor with an indirect bandgap that, 
 the valence band maximum (VBM) locates along the $\Gamma$-X line while the conduction band minimum (CBM)
 lies at $\Gamma$ as shown in Fig.~\ref{band-biaxial}. 
 To better understand the strain-depend band gap behavior, we calculated the effect of biaxial 
 strain on the band structure was presented in Fig.~\ref{band-biaxial}. Note that, irreducible Brillouin Zone 
 of GaS under biaxial strain remains unchanged, as the hexagonal symmetry is saved.  
 With tensile biaxial strain, the energy of the CBM at $\Gamma$ reduces rapidly. 
 But the reduction of energy of VBM as labeled $\Lambda$ is slower than up to +6$\%$. 
 When more strain is added, the energy of $\Lambda$ (at VBM) increases. 
 The result indicates that when biaxial tensile strain is applied the CBM and VBM always locate 
 at $\Gamma$ and $\Lambda$ point, which moves close to the Fermi level with an increase of 
 tensile strain and suggesting a stronger indirect band gap. 
 
  \begin{figure*}[ht]
\centering
\includegraphics*[scale=16.75]{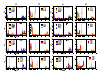}
\caption{The projected density of states of GaS monolayer under  
(a) biaxial tensile strain and (b) biaxial compressive strain. The panels a-c and b-d refer to the Ga and S atoms,
respectively. The vertical dashed lines indicate the Fermi level. The Fermi level is aligned at zero.}
\label{pdos-biaxial}
\end{figure*}

\begin{figure*}[ht]
\centering
\includegraphics*[scale=25]{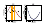}
\caption{(a) Band gap variation of GaS monolayer versus biaxial compressive strain and tensile strain in the range of $\varepsilon$=-10$\%$ to $\varepsilon$=10$\%$.
      The various gaps are indicated by the location in the irreducible Brillouin Zone were the VBM and CBM occur.
       (b) Variation of $\Delta{E}$ between the
strain-free and strained GaS monolayer as function of biaxial strain}
\label{band-gap-variation}
\end{figure*}

 Fig.~\ref{band-gap-variation} shows variation 
 of band gap as a function of biaxial strain. It can be found that, the band gap energy of GaS 
 monolayer decrease approximately linearly with increasing tensile strain. As the more tensile 
 strain applied up to $\varepsilon$=15$\%$, the band gap vanishes and consequently, 
 the monolayer becomes metallic with band crossing the Fermi level at the $\Gamma$ point.
 On the other hand, when a compressive strain of $\varepsilon$=-2$\%$ was applied to the structure, 
 the VBM remaining at $\Lambda$ while the CBM shifted to M. 
 When $\varepsilon$=-4$\%$, the CBM of monolayer switches from M to K, while keeps the VBM unchanged. 
 It is obvious that the band at A point gradually moves upward, consequently, at $\varepsilon$=-6$\%$ VBM 
 changes to $\Gamma$ point. At the same time, the B (CBM) point energy downward under compressive strain.
 As the value of the compressive strain increases similar behavior occurs with
 the VBM and CBM. But the energy shift of B point is more prominent than A point. 
 Moreover, except within $\varepsilon$=-2$\%$ and $\varepsilon$=-4$\%$ of compressive strain, the band gap
 generally reduced as the strain increase. (shown on the Fig.~\ref{band-gap-variation}). 
 It is clear that the indirect band gap can not be transformed to be direct band gap by applying biaxial strain in the 
 range of $\varepsilon$=-10$\%$ to $\varepsilon$=10$\%$. 

 To understand the bandgap engineering via biaxial strain, the projected density of states 
 (PDOSs) were calculated for the GaS monolayer. 
 It can be found that from Fig.~\ref{pdos-biaxial}, the electronic
 states of GaS monolayer near the CBM and VBM contributed mainly from $p_z$ and $s$ 
 orbitals of the Gallium atoms and the $p_y$ orbital of the Sulfur atoms. 
 It is clear that the S $p_y$ orbital plays a very important role in the VBM. 
 As the tensile strain applied, the CBM and the VBM shift toward the Fermi level. 
 At $\varepsilon$=4$\%$, the S $p_y$ orbital shifts back and VBM is replaced by S $p_z$ orbital.
 On the other hand, when a compressive strain was applied of $\varepsilon$=-2$\%$ and 
 $\varepsilon$=-4$\%$, the CBM and VBM of sulfur atoms are not contributed 
 by the same electric components as that of pristine one. 
 As shown in Fig.~\ref{pdos-biaxial} 
 at $\varepsilon$=-2$\%$ and $\varepsilon$=-4$\%$, the CBM and VBM of sulfur atoms are
 mainly contributed by $p_z$ orbital. 
 At the same time, the electronic component of VBM 
 of Ga is not changed and it is still contributed by Ga $s$ and Ga $p_z$ 
 orbitals, while the electronic components of CBM of Ga are changed and contributed by 
 Ga $p_z$ and $p_y$ orbitals. 
 The main changing in the electronic contribution of CBM and VBM of
 sulfur atoms, leading to a band gap enhancement for GaS monolayer at $\varepsilon$=-2$\%$ and 
 $\varepsilon$=-4$\%$. As the compression increases further, above $\varepsilon$=-4$\%$ the S 
 and Ga $p_y$ orbitals shift toward the Fermi level and the band gap decreases again.

 \subsection{Uniaxial strain}
 
 We have also applied uniaxial strain on rectangular GaS monolayer. Under uniaxial strain, 
 the original crystal symmetry is broken, thus the K-points become no longer equivalent as before.
 We define the $\varepsilon_x$ and $\varepsilon_y$ along the zigzag and armchair directions.
 In the band structure of the pristine GaS monolayer, the VBM location between $\Gamma$-X and the 
 CBM location at the $\Gamma$ point results in indirect bandgap (see Fig.~\ref{band-uniaxial}). 
 
 Under zigzag compressive strain, the VBM is retained 
 between $\Gamma$-X, as the value of the compressive strain increase, the energy of A point gradually 
 increases. Beyond $\varepsilon$=-6x$\%$, A has a highest energy and thus represent the VBM. 
 With an increase compression, the dispersion relation of the CBM at $\Gamma$ gradually deform, 
 therefore making the pristine CBM point at $\Gamma$ move to another energy point between $\Gamma$-X, 
 above $\varepsilon$=-4x$\%$. 
 And the indirect semiconductor character is retained. As the zigzag 
 tensile uniaxial strain applied, the band gap retain indirect. 
 With continuous increase of the strain, 
 the band gap decreases. 
 As can be found, the CBM and VBM remain at $\Gamma$ and 
 between $\Gamma$-X, respectively.\\
  
\begin{figure}[!ht]
\centering
\includegraphics*[scale=15.5]{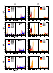}
\caption
{The PDOS on GaS monolayer under uniaxial compressive strain along y (armchair) direction. The panels a and b refer to Ga and S atoms, respectively.}
\label{pdos-uniaxial}
\end{figure}

 Under the tensile strain along the y-direction (armchair) keeping the VBM unchanged, while 
 the energy of the CBM at $\Gamma$, decreases rapidly.
 Thereby strengthening the indirect
 band gap quickly. 
 
 As the uniaxial compressive strain is taken along y direction, 
 the CBM always locates at the $\Gamma$ point with experiences approximately 0.7 eV
 downward shift up to $\varepsilon$=-10y$\%$. At the same time, as the more compression 
 was applied, the energy of a sub-VB (labelled B) raises rapidly. 
 At $\varepsilon$=-10y$\%$, the energy of B at $\Gamma$ point has a higher energy than C 
 and becomes the VBM. Therefore showing a direct band gap at $\Gamma$.\\
 
 In summary, the biaxial strain can not change the indirect band gap to direct,  
 Nevertheless, 
 has more influence on electronic band structure.  
  But the armchair compressive uniaxial strain can reduce the band gap and leading to 
 indirect-direct transition at $\varepsilon$=-10y$\%$ strain. For more details, the PDOS were 
 calculated as shown in Fig.~\ref{pdos-uniaxial}. By analyzing the PDOS under armchair strain 
 found that the role of $p_y$ orbital gradually increase for both Ga and S atoms. 
 With the continuous increase of the strain at $\varepsilon$=-10y$\%$ (where indirect-direct transition happens) 
 the CBM and VBM are mainly contributed by S $p_y$ $p_z$ Ga $s$ $p_y$ orbitals.
 Therefore, the band gap transition occurs at $\Gamma$ point caused by the increasing 
 contribution of $p_y$ orbital.

  \section{Charge transfer}
 
 In this section, we examine the charge transfer between Ga and S atoms By Bader analysis \cite{henkelman2006fast}. 
 The depletion of charge on Ga atom is 0.82e in pristine GaS monolayer, whereas excess charge
 on S atoms that proves the ionicity in Ga-S bonding.
 The charge transfer from S to Ga increases with increasing compressive biaxial strain. 
 Thus, we conclude that the Ga-S bonds become less ionic and mix ionic-covalent in comparison
 with equilibrium structure. As the large tensile strain applied, 
 (above $\varepsilon$=6$\%$) the charge of sulfur and Gallium come back to near isolate atom 
 because of the distance between S and Ga increases.

 \section{Mechanical properties}

\begin{figure*}[ht]
\centering
\includegraphics*[scale=30]{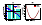}
\caption{(a) The applied axial strains in the zigzag and armchair direction, respectively. 
(b) The evolution of strain energy of GaS monolayer under uniaxial strain along
the x (black) and y (dashed red) directons}
\label{strain energy}
\end{figure*}

 The strain in the x or y-direction 
 and the transverse strain in the y or x-direction are shown in Fig.~\ref{strain energy}. 
 for the GaS monolayer Poission's ratio:
 \begin{align}
 \nu=d\epsilon_{transverse}/d\epsilon_{axial}  
 \end{align}
 
are calculated by fitting the figure of transverse strain response. The obtained results are 
0.23 and 0.24 under zigzag and armchair uniaxial strain, respectively. These two close
value of Poission's ratio indicate the isotropic nature of the mechanical properties of 
the GaS monolayer. In another word, we understand the mechanical properties are independent of direction.
While phosphorene has anisotropic nature of mechanical properties with 0.24 and 0.7 
value for Poission's ration in zigzag and armchair direction \cite{peng2014strain}.\\

In the Fig.~\ref{strain energy}, we plotted strain energy for two directions 
(zigzag and armchair) $E_{strain}$ as:
 \begin{align}
 E_{strian}= E_T(\varepsilon)-E_T(\varepsilon=0)  
 \end{align}
 where $E_T(\varepsilon)$ and $E_T(\varepsilon=0)$ are the total energies of the GaS monolayer with strain and strain-free, 
 respectively. 
 It is clear that the near 
 degeneracy of strain energy in zigzag and armchair directions in the range 
 of $\varepsilon$=-10$\%$ to $\varepsilon$=10$\%$, demonstrate that, for 
 applying strain in either zigzag or armchair directions need equal energy.
 In addition, confirm the isotropic nature of mechanical properties. In the end, 
 we calculated the energy difference ($\Delta{E}$)
 between strained and strain-free GaS monolayer as the function of biaxial strain. 
 
 The $\Delta{E}$ increases monotonously (Fig.\ref{band-gap-variation}) 
 for both tensile and compressive strains, shows that the system is in the range of the elastic deformation. 
 Therefore, after removing applied strain, the structure releases back to the original structure which is 
 important for sensor applications.
 
 \section{Nanoribbon}
 
\begin{figure}[!ht]
\centering
\includegraphics*[scale=21]{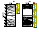}
\caption{(a) The calculated band structure of H-saturated 10-AGaSNR (Left) and the relaxed H-saturated structure of 10-AGaSNR (Right). (b) The calculated band structure of
H-saturated 4-ZGaSNR (Left) and the relaxed H-saturated structure of 4-ZGaSNR (Right). The blue, yellow and red spheres represent Ga, S and H atoms. The Fermi level is aligned at zero.}
\label{NR}
\end{figure}

 We have also investigated the electronic properties of bare and hydrogen saturated zigzag/armchair 
 GaS nanoribbons to find indirect-direct band gap transition. 
 First, we calculated the band structure of bare-edge armchair
 GaS nanoribbon with a width ranging from $N_a=4$ to $N_a=10$. 
 We found that all armchair nanoribbons are indirect band gap semiconductor. 
 with E$_g$ values of approximately 1.47 eV, independent of the width, 
 which is smaller than the monolayer band gap(2.45 eV) 
 and can not be interpreted by the well-known quantum confinement effect.  
The new flat energy level at both conduction and valence band edges 
came into being via edge atoms of nanoribbons presence.
engineering the band structure, both edges are passivated with hydrogen atoms (each edge 
 Ga and S atoms are terminated by one H atom). Fig.~\ref{NR} depicts the band structure for 
 passivated armchair GaS nanoribbons. 
 It reveals that the AGaSNR remain semiconductor after 
 hydrogen saturation and keep semiconducting independent with the ribbon width. 
 After passivation, the band gap experiences a big jump from 1.47eV to 2.48 eV for all 
 nanoribbons, approximately. As the width of the nanoribbons increase, the band gap decrease, slowly. 
 At the same time, the band at $\Gamma$ point moves downward (not shown here) and become CBM in 10-AGaSNR. 
 Consequently, the band structure undergoes an indirect to
 direct transitions in the 10-AGaSNR passivated. 
 In addition, the energy bands around the Fermi level are rather flat for all 
 the investigated nanoribbons Which suggest a large effective 
 mass and improvement in their transport properties.\\
 
 Also, we examine the zigzag GaS nanoribbons with a width ranging from $N_z$=4 to $N_z$=10. 
 The results indicated that, in contrast to armchair GaS nanoribbons, the zigzag structure are
 metallic and their metallic behavior is independent of ribbon width. As can be found that 
 from Fig.~\ref{NR}, when saturating the zigzag nanoribbons edge with hydrogen, all of 
 the nanoribbons retain metallic behavior, except 4-ZGaSNR that become semiconductor 
 with a small indirect band gap of 0.16 eV and direct band gap of 0.15 eV. 
 It caused by edge effects in 4-ZGaSNR that influences each other.
 
  \section{CONCLUSION}
  
 In conclusion, the first-principles DFT calculations introduce some special circumstance 
 in which GaS structures experience indirect-direct band gap transition.
Applying the mechanical deformation via biaxial and uniaxial strain, 
not only the Ga-S bond length, but also the angle
between S-Ga-S atoms have a significant impact on the band 
structure of GaS monolayer. 
 We show that under biaxial strain in the range of $\varepsilon$=-10$\%$ 
 to 10$\%$, the indirect to direct transition of GaS monolayer did not observe. 
 The band gap always decreases under strain except at $\varepsilon$=-2$\%$ 
 and $\varepsilon$=-4$\%$ because the $p_z$ orbital shift towards the Fermi 
 level and above  $\varepsilon$=-4$\%$, replaced by $p_y$ again and the GaS 
 monolayer exhibits a semiconductor to metal transition at $\varepsilon$=+15$\%$. 
 Under uniaxial compressive strain along y (armchair) direction, the upward shift of 
 sub-VBM is more than VBM and become VBM at $\Gamma$ point and leading to 
 crossover from indirect to direct band gap at $\varepsilon$=-10$\%$ due to $p_y$ orbital 
 move toward the Fermi level in both Ga and S atoms.\\
 The Poission's ratio along armchair and zigzag directions 
 has isotropic nature and we can tailor GaS band gap with uniaxial $\varepsilon$=-10$\%$ strain
 without concern to mechanical anisotropy in contrast to phosphorene that posses anisotropic nature of mechanical 
  properties.\\
 For nanoribbon structures, we find that both bare and hydrogen saturated 
 AGaSNR are a semiconductor, while the ZGaSNRs are metallic. The  
 semiconductor behavior of the AGaSNR is independent of the ribbon width.
The band gaps of H-saturated AGaSNR increases in comparison to 
bare-edge and the size of band gap do not vary significantly with ribbon width. 
The indirect to direct band gap transition observed in 10-AGaSNR. 
 when the armchair ribbon edges are saturated with hydrogen atoms. 
 The 4-ZGaSNR exhibits an indirect band gap of
 0.16eV and the direct band gap of 0.15eV. Another H-saturated $N_z$-ZGaSNR with 
 $N_z$=6 - 8 - 10 retain metallic behavior.
 The improvement towards controlling the band structure and optoelectronic 
properties of GaS monolayer achieved with strain engineering.\\

\section{ACKNOWLEDGMENTS}

We would like to acknowledge the Isfahan University of Technology. The authors gratefully acknowledge 
the Sheikh Bahaei National High Performance Computing Center (SBNHPCC) for providing computing 
facilities and time. SBNHPCC is supported by scientific and technological department of presidential 
office and Isfahan University of Technology (IUT).

\bibliography{ref}

\begin{thebibliography}{26}%
\makeatletter
\providecommand \@ifxundefined [1]{%
 \@ifx{#1\undefined}
}%
\providecommand \@ifnum [1]{%
 \ifnum #1\expandafter \@firstoftwo
 \else \expandafter \@secondoftwo
 \fi
}%
\providecommand \@ifx [1]{%
 \ifx #1\expandafter \@firstoftwo
 \else \expandafter \@secondoftwo
 \fi
}%
\providecommand \natexlab [1]{#1}%
\providecommand \enquote  [1]{``#1''}%
\providecommand \bibnamefont  [1]{#1}%
\providecommand \bibfnamefont [1]{#1}%
\providecommand \citenamefont [1]{#1}%
\providecommand \href@noop [0]{\@secondoftwo}%
\providecommand \href [0]{\begingroup \@sanitize@url \@href}%
\providecommand \@href[1]{\@@startlink{#1}\@@href}%
\providecommand \@@href[1]{\endgroup#1\@@endlink}%
\providecommand \@sanitize@url [0]{\catcode `\\12\catcode `\$12\catcode
  `\&12\catcode `\#12\catcode `\^12\catcode `\_12\catcode `\%12\relax}%
\providecommand \@@startlink[1]{}%
\providecommand \@@endlink[0]{}%
\providecommand \url  [0]{\begingroup\@sanitize@url \@url }%
\providecommand \@url [1]{\endgroup\@href {#1}{\urlprefix }}%
\providecommand \urlprefix  [0]{URL }%
\providecommand \Eprint [0]{\href }%
\providecommand \doibase [0]{http://dx.doi.org/}%
\providecommand \selectlanguage [0]{\@gobble}%
\providecommand \bibinfo  [0]{\@secondoftwo}%
\providecommand \bibfield  [0]{\@secondoftwo}%
\providecommand \translation [1]{[#1]}%
\providecommand \BibitemOpen [0]{}%
\providecommand \bibitemStop [0]{}%
\providecommand \bibitemNoStop [0]{.\EOS\space}%
\providecommand \EOS [0]{\spacefactor3000\relax}%
\providecommand \BibitemShut  [1]{\csname bibitem#1\endcsname}%
\let\auto@bib@innerbib\@empty
\bibitem [{\citenamefont {Gupta}\ \emph {et~al.}(2015)\citenamefont {Gupta},
  \citenamefont {Sakthivel},\ and\ \citenamefont {Seal}}]{gupta2015recent}%
  \BibitemOpen
  \bibfield  {author} {\bibinfo {author} {\bibfnamefont {A.}~\bibnamefont
  {Gupta}}, \bibinfo {author} {\bibfnamefont {T.}~\bibnamefont {Sakthivel}}, \
  and\ \bibinfo {author} {\bibfnamefont {S.}~\bibnamefont {Seal}},\ }\href@noop
  {} {\bibfield  {journal} {\bibinfo  {journal} {Progress in Materials
  Science}\ }\textbf {\bibinfo {volume} {73}},\ \bibinfo {pages} {44} (\bibinfo
  {year} {2015})}\BibitemShut {NoStop}%
\bibitem [{\citenamefont {Radisavljevic}\ \emph {et~al.}(2011)\citenamefont
  {Radisavljevic}, \citenamefont {Radenovic}, \citenamefont {Brivio},
  \citenamefont {Giacometti},\ and\ \citenamefont
  {Kis}}]{radisavljevic2011single}%
  \BibitemOpen
  \bibfield  {author} {\bibinfo {author} {\bibfnamefont {B.}~\bibnamefont
  {Radisavljevic}}, \bibinfo {author} {\bibfnamefont {A.}~\bibnamefont
  {Radenovic}}, \bibinfo {author} {\bibfnamefont {J.}~\bibnamefont {Brivio}},
  \bibinfo {author} {\bibfnamefont {i.~V.}\ \bibnamefont {Giacometti}}, \ and\
  \bibinfo {author} {\bibfnamefont {A.}~\bibnamefont {Kis}},\ }\href@noop {}
  {\bibfield  {journal} {\bibinfo  {journal} {Nature nanotechnology}\ }\textbf
  {\bibinfo {volume} {6}},\ \bibinfo {pages} {147} (\bibinfo {year}
  {2011})}\BibitemShut {NoStop}%
\bibitem [{\citenamefont {Late}\ \emph
  {et~al.}(2012{\natexlab{a}})\citenamefont {Late}, \citenamefont {Liu},
  \citenamefont {Matte}, \citenamefont {Rao},\ and\ \citenamefont
  {Dravid}}]{late2012rapid}%
  \BibitemOpen
  \bibfield  {author} {\bibinfo {author} {\bibfnamefont {D.~J.}\ \bibnamefont
  {Late}}, \bibinfo {author} {\bibfnamefont {B.}~\bibnamefont {Liu}}, \bibinfo
  {author} {\bibfnamefont {H.~R.}\ \bibnamefont {Matte}}, \bibinfo {author}
  {\bibfnamefont {C.}~\bibnamefont {Rao}}, \ and\ \bibinfo {author}
  {\bibfnamefont {V.~P.}\ \bibnamefont {Dravid}},\ }\href@noop {} {\bibfield
  {journal} {\bibinfo  {journal} {Advanced Functional Materials}\ }\textbf
  {\bibinfo {volume} {22}},\ \bibinfo {pages} {1894} (\bibinfo {year}
  {2012}{\natexlab{a}})}\BibitemShut {NoStop}%
\bibitem [{\citenamefont {Aono}\ \emph {et~al.}(1993)\citenamefont {Aono},
  \citenamefont {Kase},\ and\ \citenamefont {Kinoshita}}]{aono1993near}%
  \BibitemOpen
  \bibfield  {author} {\bibinfo {author} {\bibfnamefont {T.}~\bibnamefont
  {Aono}}, \bibinfo {author} {\bibfnamefont {K.}~\bibnamefont {Kase}}, \ and\
  \bibinfo {author} {\bibfnamefont {A.}~\bibnamefont {Kinoshita}},\ }\href@noop
  {} {\bibfield  {journal} {\bibinfo  {journal} {Journal of applied physics}\
  }\textbf {\bibinfo {volume} {74}},\ \bibinfo {pages} {2818} (\bibinfo {year}
  {1993})}\BibitemShut {NoStop}%
\bibitem [{\citenamefont {Shen}\ \emph {et~al.}(2009)\citenamefont {Shen},
  \citenamefont {Chen}, \citenamefont {Chen},\ and\ \citenamefont
  {Zhou}}]{shen2009vapor}%
  \BibitemOpen
  \bibfield  {author} {\bibinfo {author} {\bibfnamefont {G.}~\bibnamefont
  {Shen}}, \bibinfo {author} {\bibfnamefont {D.}~\bibnamefont {Chen}}, \bibinfo
  {author} {\bibfnamefont {P.-C.}\ \bibnamefont {Chen}}, \ and\ \bibinfo
  {author} {\bibfnamefont {C.}~\bibnamefont {Zhou}},\ }\href@noop {} {\bibfield
   {journal} {\bibinfo  {journal} {Acs Nano}\ }\textbf {\bibinfo {volume}
  {3}},\ \bibinfo {pages} {1115} (\bibinfo {year} {2009})}\BibitemShut
  {NoStop}%
\bibitem [{\citenamefont {Hu}\ \emph {et~al.}(2013)\citenamefont {Hu},
  \citenamefont {Wang}, \citenamefont {Yoon}, \citenamefont {Zhang},
  \citenamefont {Feng}, \citenamefont {Wang}, \citenamefont {Wen},
  \citenamefont {Idrobo}, \citenamefont {Miyamoto}, \citenamefont {Geohegan}
  \emph {et~al.}}]{hu2013highly}%
  \BibitemOpen
  \bibfield  {author} {\bibinfo {author} {\bibfnamefont {P.}~\bibnamefont
  {Hu}}, \bibinfo {author} {\bibfnamefont {L.}~\bibnamefont {Wang}}, \bibinfo
  {author} {\bibfnamefont {M.}~\bibnamefont {Yoon}}, \bibinfo {author}
  {\bibfnamefont {J.}~\bibnamefont {Zhang}}, \bibinfo {author} {\bibfnamefont
  {W.}~\bibnamefont {Feng}}, \bibinfo {author} {\bibfnamefont {X.}~\bibnamefont
  {Wang}}, \bibinfo {author} {\bibfnamefont {Z.}~\bibnamefont {Wen}}, \bibinfo
  {author} {\bibfnamefont {J.~C.}\ \bibnamefont {Idrobo}}, \bibinfo {author}
  {\bibfnamefont {Y.}~\bibnamefont {Miyamoto}}, \bibinfo {author}
  {\bibfnamefont {D.~B.}\ \bibnamefont {Geohegan}},  \emph {et~al.},\
  }\href@noop {} {\bibfield  {journal} {\bibinfo  {journal} {Nano letters}\
  }\textbf {\bibinfo {volume} {13}},\ \bibinfo {pages} {1649} (\bibinfo {year}
  {2013})}\BibitemShut {NoStop}%
\bibitem [{\citenamefont {Late}\ \emph
  {et~al.}(2012{\natexlab{b}})\citenamefont {Late}, \citenamefont {Liu},
  \citenamefont {Luo}, \citenamefont {Yan}, \citenamefont {Matte},
  \citenamefont {Grayson}, \citenamefont {Rao},\ and\ \citenamefont
  {Dravid}}]{late2012gas}%
  \BibitemOpen
  \bibfield  {author} {\bibinfo {author} {\bibfnamefont {D.~J.}\ \bibnamefont
  {Late}}, \bibinfo {author} {\bibfnamefont {B.}~\bibnamefont {Liu}}, \bibinfo
  {author} {\bibfnamefont {J.}~\bibnamefont {Luo}}, \bibinfo {author}
  {\bibfnamefont {A.}~\bibnamefont {Yan}}, \bibinfo {author} {\bibfnamefont
  {H.~R.}\ \bibnamefont {Matte}}, \bibinfo {author} {\bibfnamefont
  {M.}~\bibnamefont {Grayson}}, \bibinfo {author} {\bibfnamefont
  {C.}~\bibnamefont {Rao}}, \ and\ \bibinfo {author} {\bibfnamefont {V.~P.}\
  \bibnamefont {Dravid}},\ }\href@noop {} {\bibfield  {journal} {\bibinfo
  {journal} {Advanced Materials}\ }\textbf {\bibinfo {volume} {24}},\ \bibinfo
  {pages} {3549} (\bibinfo {year} {2012}{\natexlab{b}})}\BibitemShut {NoStop}%
\bibitem [{\citenamefont {Li}\ and\ \citenamefont
  {Li}(2015)}]{li2015piezoelectricity}%
  \BibitemOpen
  \bibfield  {author} {\bibinfo {author} {\bibfnamefont {W.}~\bibnamefont
  {Li}}\ and\ \bibinfo {author} {\bibfnamefont {J.}~\bibnamefont {Li}},\
  }\href@noop {} {\bibfield  {journal} {\bibinfo  {journal} {Nano Research}\
  }\textbf {\bibinfo {volume} {8}},\ \bibinfo {pages} {3796} (\bibinfo {year}
  {2015})}\BibitemShut {NoStop}%
\bibitem [{\citenamefont {Topsakal}\ \emph {et~al.}(2010)\citenamefont
  {Topsakal}, \citenamefont {Cahangirov},\ and\ \citenamefont
  {Ciraci}}]{topsakal2010response}%
  \BibitemOpen
  \bibfield  {author} {\bibinfo {author} {\bibfnamefont {M.}~\bibnamefont
  {Topsakal}}, \bibinfo {author} {\bibfnamefont {S.}~\bibnamefont
  {Cahangirov}}, \ and\ \bibinfo {author} {\bibfnamefont {S.}~\bibnamefont
  {Ciraci}},\ }\href@noop {} {\bibfield  {journal} {\bibinfo  {journal}
  {Applied Physics Letters}\ }\textbf {\bibinfo {volume} {96}},\ \bibinfo
  {pages} {091912} (\bibinfo {year} {2010})}\BibitemShut {NoStop}%
\bibitem [{\citenamefont {Choi}\ \emph {et~al.}(2010)\citenamefont {Choi},
  \citenamefont {Jhi},\ and\ \citenamefont {Son}}]{choi2010effects}%
  \BibitemOpen
  \bibfield  {author} {\bibinfo {author} {\bibfnamefont {S.-M.}\ \bibnamefont
  {Choi}}, \bibinfo {author} {\bibfnamefont {S.-H.}\ \bibnamefont {Jhi}}, \
  and\ \bibinfo {author} {\bibfnamefont {Y.-W.}\ \bibnamefont {Son}},\
  }\href@noop {} {\bibfield  {journal} {\bibinfo  {journal} {Physical Review
  B}\ }\textbf {\bibinfo {volume} {81}},\ \bibinfo {pages} {081407} (\bibinfo
  {year} {2010})}\BibitemShut {NoStop}%
\bibitem [{\citenamefont {Peelaers}\ and\ \citenamefont {Van~de
  Walle}(2012)}]{peelaers2012effects}%
  \BibitemOpen
  \bibfield  {author} {\bibinfo {author} {\bibfnamefont {H.}~\bibnamefont
  {Peelaers}}\ and\ \bibinfo {author} {\bibfnamefont {C.~G.}\ \bibnamefont
  {Van~de Walle}},\ }\href@noop {} {\bibfield  {journal} {\bibinfo  {journal}
  {Physical Review B}\ }\textbf {\bibinfo {volume} {86}},\ \bibinfo {pages}
  {241401} (\bibinfo {year} {2012})}\BibitemShut {NoStop}%
\bibitem [{\citenamefont {Scalise}\ \emph {et~al.}(2012)\citenamefont
  {Scalise}, \citenamefont {Houssa}, \citenamefont {Pourtois}, \citenamefont
  {Afanas’ev},\ and\ \citenamefont {Stesmans}}]{scalise2012strain}%
  \BibitemOpen
  \bibfield  {author} {\bibinfo {author} {\bibfnamefont {E.}~\bibnamefont
  {Scalise}}, \bibinfo {author} {\bibfnamefont {M.}~\bibnamefont {Houssa}},
  \bibinfo {author} {\bibfnamefont {G.}~\bibnamefont {Pourtois}}, \bibinfo
  {author} {\bibfnamefont {V.}~\bibnamefont {Afanas’ev}}, \ and\ \bibinfo
  {author} {\bibfnamefont {A.}~\bibnamefont {Stesmans}},\ }\href@noop {}
  {\bibfield  {journal} {\bibinfo  {journal} {Nano Research}\ }\textbf
  {\bibinfo {volume} {5}},\ \bibinfo {pages} {43} (\bibinfo {year}
  {2012})}\BibitemShut {NoStop}%
\bibitem [{\citenamefont {Ghorbani-Asl}\ \emph {et~al.}(2013)\citenamefont
  {Ghorbani-Asl}, \citenamefont {Borini}, \citenamefont {Kuc},\ and\
  \citenamefont {Heine}}]{ghorbani2013strain}%
  \BibitemOpen
  \bibfield  {author} {\bibinfo {author} {\bibfnamefont {M.}~\bibnamefont
  {Ghorbani-Asl}}, \bibinfo {author} {\bibfnamefont {S.}~\bibnamefont
  {Borini}}, \bibinfo {author} {\bibfnamefont {A.}~\bibnamefont {Kuc}}, \ and\
  \bibinfo {author} {\bibfnamefont {T.}~\bibnamefont {Heine}},\ }\href@noop {}
  {\bibfield  {journal} {\bibinfo  {journal} {Physical Review B}\ }\textbf
  {\bibinfo {volume} {87}},\ \bibinfo {pages} {235434} (\bibinfo {year}
  {2013})}\BibitemShut {NoStop}%
\bibitem [{\citenamefont {Sun}\ \emph {et~al.}(2016)\citenamefont {Sun},
  \citenamefont {Wang},\ and\ \citenamefont {Shuai}}]{sun2016indirect}%
  \BibitemOpen
  \bibfield  {author} {\bibinfo {author} {\bibfnamefont {Y.}~\bibnamefont
  {Sun}}, \bibinfo {author} {\bibfnamefont {D.}~\bibnamefont {Wang}}, \ and\
  \bibinfo {author} {\bibfnamefont {Z.}~\bibnamefont {Shuai}},\ }\href@noop {}
  {\bibfield  {journal} {\bibinfo  {journal} {The Journal of Physical Chemistry
  C}\ }\textbf {\bibinfo {volume} {120}},\ \bibinfo {pages} {21866} (\bibinfo
  {year} {2016})}\BibitemShut {NoStop}%
\bibitem [{\citenamefont {Wang}\ \emph {et~al.}(2015)\citenamefont {Wang},
  \citenamefont {Cong}, \citenamefont {Yang}, \citenamefont {Shang},
  \citenamefont {Peimyoo}, \citenamefont {Chen}, \citenamefont {Kang},
  \citenamefont {Wang}, \citenamefont {Huang},\ and\ \citenamefont
  {Yu}}]{wang2015strain}%
  \BibitemOpen
  \bibfield  {author} {\bibinfo {author} {\bibfnamefont {Y.}~\bibnamefont
  {Wang}}, \bibinfo {author} {\bibfnamefont {C.}~\bibnamefont {Cong}}, \bibinfo
  {author} {\bibfnamefont {W.}~\bibnamefont {Yang}}, \bibinfo {author}
  {\bibfnamefont {J.}~\bibnamefont {Shang}}, \bibinfo {author} {\bibfnamefont
  {N.}~\bibnamefont {Peimyoo}}, \bibinfo {author} {\bibfnamefont
  {Y.}~\bibnamefont {Chen}}, \bibinfo {author} {\bibfnamefont {J.}~\bibnamefont
  {Kang}}, \bibinfo {author} {\bibfnamefont {J.}~\bibnamefont {Wang}}, \bibinfo
  {author} {\bibfnamefont {W.}~\bibnamefont {Huang}}, \ and\ \bibinfo {author}
  {\bibfnamefont {T.}~\bibnamefont {Yu}},\ }\href@noop {} {\bibfield  {journal}
  {\bibinfo  {journal} {Nano Research}\ }\textbf {\bibinfo {volume} {8}},\
  \bibinfo {pages} {2562} (\bibinfo {year} {2015})}\BibitemShut {NoStop}%
\bibitem [{\citenamefont {Ni}\ \emph {et~al.}(2008)\citenamefont {Ni},
  \citenamefont {Yu}, \citenamefont {Lu}, \citenamefont {Wang}, \citenamefont
  {Feng},\ and\ \citenamefont {Shen}}]{ni2008uniaxial}%
  \BibitemOpen
  \bibfield  {author} {\bibinfo {author} {\bibfnamefont {Z.~H.}\ \bibnamefont
  {Ni}}, \bibinfo {author} {\bibfnamefont {T.}~\bibnamefont {Yu}}, \bibinfo
  {author} {\bibfnamefont {Y.~H.}\ \bibnamefont {Lu}}, \bibinfo {author}
  {\bibfnamefont {Y.~Y.}\ \bibnamefont {Wang}}, \bibinfo {author}
  {\bibfnamefont {Y.~P.}\ \bibnamefont {Feng}}, \ and\ \bibinfo {author}
  {\bibfnamefont {Z.~X.}\ \bibnamefont {Shen}},\ }\href@noop {} {\bibfield
  {journal} {\bibinfo  {journal} {ACS nano}\ }\textbf {\bibinfo {volume} {2}},\
  \bibinfo {pages} {2301} (\bibinfo {year} {2008})}\BibitemShut {NoStop}%
\bibitem [{\citenamefont {Huang}\ \emph {et~al.}(2011)\citenamefont {Huang},
  \citenamefont {Yu},\ and\ \citenamefont {Wei}}]{huang2011strain}%
  \BibitemOpen
  \bibfield  {author} {\bibinfo {author} {\bibfnamefont {B.}~\bibnamefont
  {Huang}}, \bibinfo {author} {\bibfnamefont {J.}~\bibnamefont {Yu}}, \ and\
  \bibinfo {author} {\bibfnamefont {S.-H.}\ \bibnamefont {Wei}},\ }\href@noop
  {} {\bibfield  {journal} {\bibinfo  {journal} {Physical Review B}\ }\textbf
  {\bibinfo {volume} {84}},\ \bibinfo {pages} {075415} (\bibinfo {year}
  {2011})}\BibitemShut {NoStop}%
\bibitem [{\citenamefont {Shi}\ \emph {et~al.}(2013)\citenamefont {Shi},
  \citenamefont {Pan}, \citenamefont {Zhang},\ and\ \citenamefont
  {Yakobson}}]{shi2013quasiparticle}%
  \BibitemOpen
  \bibfield  {author} {\bibinfo {author} {\bibfnamefont {H.}~\bibnamefont
  {Shi}}, \bibinfo {author} {\bibfnamefont {H.}~\bibnamefont {Pan}}, \bibinfo
  {author} {\bibfnamefont {Y.-W.}\ \bibnamefont {Zhang}}, \ and\ \bibinfo
  {author} {\bibfnamefont {B.~I.}\ \bibnamefont {Yakobson}},\ }\href@noop {}
  {\bibfield  {journal} {\bibinfo  {journal} {Physical Review B}\ }\textbf
  {\bibinfo {volume} {87}},\ \bibinfo {pages} {155304} (\bibinfo {year}
  {2013})}\BibitemShut {NoStop}%
\bibitem [{\citenamefont {Bl{\"o}chl}(1994)}]{blochl1994projector}%
  \BibitemOpen
  \bibfield  {author} {\bibinfo {author} {\bibfnamefont {P.~E.}\ \bibnamefont
  {Bl{\"o}chl}},\ }\href@noop {} {\bibfield  {journal} {\bibinfo  {journal}
  {Physical review B}\ }\textbf {\bibinfo {volume} {50}},\ \bibinfo {pages}
  {17953} (\bibinfo {year} {1994})}\BibitemShut {NoStop}%
\bibitem [{\citenamefont {Perdew}(1996)}]{perdew1996jp}%
  \BibitemOpen
  \bibfield  {author} {\bibinfo {author} {\bibfnamefont {J.~P.}\ \bibnamefont
  {Perdew}},\ }\href@noop {} {\bibfield  {journal} {\bibinfo  {journal} {Phys.
  Rev. Lett.}\ }\textbf {\bibinfo {volume} {77}},\ \bibinfo {pages} {3865}
  (\bibinfo {year} {1996})}\BibitemShut {NoStop}%
\bibitem [{\citenamefont {Giannozzi}\ \emph {et~al.}(2009)\citenamefont
  {Giannozzi}, \citenamefont {Baroni}, \citenamefont {Bonini}, \citenamefont
  {Calandra}, \citenamefont {Car}, \citenamefont {Cavazzoni}, \citenamefont
  {Ceresoli}, \citenamefont {Chiarotti}, \citenamefont {Cococcioni},
  \citenamefont {Dabo} \emph {et~al.}}]{giannozzi2009quantum}%
  \BibitemOpen
  \bibfield  {author} {\bibinfo {author} {\bibfnamefont {P.}~\bibnamefont
  {Giannozzi}}, \bibinfo {author} {\bibfnamefont {S.}~\bibnamefont {Baroni}},
  \bibinfo {author} {\bibfnamefont {N.}~\bibnamefont {Bonini}}, \bibinfo
  {author} {\bibfnamefont {M.}~\bibnamefont {Calandra}}, \bibinfo {author}
  {\bibfnamefont {R.}~\bibnamefont {Car}}, \bibinfo {author} {\bibfnamefont
  {C.}~\bibnamefont {Cavazzoni}}, \bibinfo {author} {\bibfnamefont
  {D.}~\bibnamefont {Ceresoli}}, \bibinfo {author} {\bibfnamefont {G.~L.}\
  \bibnamefont {Chiarotti}}, \bibinfo {author} {\bibfnamefont {M.}~\bibnamefont
  {Cococcioni}}, \bibinfo {author} {\bibfnamefont {I.}~\bibnamefont {Dabo}},
  \emph {et~al.},\ }\href@noop {} {\bibfield  {journal} {\bibinfo  {journal}
  {Journal of physics: Condensed matter}\ }\textbf {\bibinfo {volume} {21}},\
  \bibinfo {pages} {395502} (\bibinfo {year} {2009})}\BibitemShut {NoStop}%
\bibitem [{\citenamefont {Henkelman}\ \emph {et~al.}(2006)\citenamefont
  {Henkelman}, \citenamefont {Arnaldsson},\ and\ \citenamefont
  {J{\'o}nsson}}]{henkelman2006fast}%
  \BibitemOpen
  \bibfield  {author} {\bibinfo {author} {\bibfnamefont {G.}~\bibnamefont
  {Henkelman}}, \bibinfo {author} {\bibfnamefont {A.}~\bibnamefont
  {Arnaldsson}}, \ and\ \bibinfo {author} {\bibfnamefont {H.}~\bibnamefont
  {J{\'o}nsson}},\ }\href@noop {} {\bibfield  {journal} {\bibinfo  {journal}
  {Computational Materials Science}\ }\textbf {\bibinfo {volume} {36}},\
  \bibinfo {pages} {354} (\bibinfo {year} {2006})}\BibitemShut {NoStop}%
\bibitem [{\citenamefont {Zolyomi}\ \emph {et~al.}(2013)\citenamefont
  {Zolyomi}, \citenamefont {Drummond},\ and\ \citenamefont
  {Fal'Ko}}]{zolyomi2013band}%
  \BibitemOpen
  \bibfield  {author} {\bibinfo {author} {\bibfnamefont {V.}~\bibnamefont
  {Zolyomi}}, \bibinfo {author} {\bibfnamefont {N.}~\bibnamefont {Drummond}}, \
  and\ \bibinfo {author} {\bibfnamefont {V.}~\bibnamefont {Fal'Ko}},\
  }\href@noop {} {\bibfield  {journal} {\bibinfo  {journal} {Physical Review
  B}\ }\textbf {\bibinfo {volume} {87}},\ \bibinfo {pages} {195403} (\bibinfo
  {year} {2013})}\BibitemShut {NoStop}%
\bibitem [{\citenamefont {Huang}\ \emph {et~al.}(2015)\citenamefont {Huang},
  \citenamefont {Chen},\ and\ \citenamefont {Li}}]{huang2015effects}%
  \BibitemOpen
  \bibfield  {author} {\bibinfo {author} {\bibfnamefont {L.}~\bibnamefont
  {Huang}}, \bibinfo {author} {\bibfnamefont {Z.}~\bibnamefont {Chen}}, \ and\
  \bibinfo {author} {\bibfnamefont {J.}~\bibnamefont {Li}},\ }\href@noop {}
  {\bibfield  {journal} {\bibinfo  {journal} {RSC Advances}\ }\textbf {\bibinfo
  {volume} {5}},\ \bibinfo {pages} {5788} (\bibinfo {year} {2015})}\BibitemShut
  {NoStop}%
\bibitem [{\citenamefont {Peng}\ \emph {et~al.}(2006)\citenamefont {Peng},
  \citenamefont {Ganti}, \citenamefont {Alizadeh}, \citenamefont {Sharma},
  \citenamefont {Kumar},\ and\ \citenamefont {Nayak}}]{peng2006strain}%
  \BibitemOpen
  \bibfield  {author} {\bibinfo {author} {\bibfnamefont {X.-H.}\ \bibnamefont
  {Peng}}, \bibinfo {author} {\bibfnamefont {S.}~\bibnamefont {Ganti}},
  \bibinfo {author} {\bibfnamefont {A.}~\bibnamefont {Alizadeh}}, \bibinfo
  {author} {\bibfnamefont {P.}~\bibnamefont {Sharma}}, \bibinfo {author}
  {\bibfnamefont {S.}~\bibnamefont {Kumar}}, \ and\ \bibinfo {author}
  {\bibfnamefont {S.}~\bibnamefont {Nayak}},\ }\href@noop {} {\bibfield
  {journal} {\bibinfo  {journal} {Physical Review B}\ }\textbf {\bibinfo
  {volume} {74}},\ \bibinfo {pages} {035339} (\bibinfo {year}
  {2006})}\BibitemShut {NoStop}%
\bibitem [{\citenamefont {Peng}\ \emph {et~al.}(2014)\citenamefont {Peng},
  \citenamefont {Wei},\ and\ \citenamefont {Copple}}]{peng2014strain}%
  \BibitemOpen
  \bibfield  {author} {\bibinfo {author} {\bibfnamefont {X.}~\bibnamefont
  {Peng}}, \bibinfo {author} {\bibfnamefont {Q.}~\bibnamefont {Wei}}, \ and\
  \bibinfo {author} {\bibfnamefont {A.}~\bibnamefont {Copple}},\ }\href@noop {}
  {\bibfield  {journal} {\bibinfo  {journal} {Physical Review B}\ }\textbf
  {\bibinfo {volume} {90}},\ \bibinfo {pages} {085402} (\bibinfo {year}
  {2014})}\BibitemShut {NoStop}%
\end{thebibliography}%
\end{document}